# Capacity Optimization through Sensing Threshold Adaptation for Cognitive Radio Networks

Fotis T. Foukalas[1], George T. Karetsos[2] and Lazaros F. Merakos[1]

[1]*Dept. of Informatics & Telecommunications, National Kapodistrian University of Athens, Ilisia, Athens, Greece*

e-mail: {foukalas,merakos}@di.uoa.gr

[2]*Dept. of Information Technology &. Telecommunications, TEI of Larissa, Larissa, Greece*

e-mail: karetsos@teilar.gr

**Abstract** In this paper we propose the capacity optimization over sensing threshold for sensing-based cognitive radio networks. The objective function of the proposed optimization is to maximize the capacity at the secondary user subject to the constraints on the transmit power and the sensing threshold in order to protect the primary user. The defined optimization problem is a convex optimization over the transmit power and the sensing threshold where the concavity on sensing threshold is proved. The problem is solved by using Lagrange duality decomposition method in conjunction with a subgradient iterative algorithm and the numerical results show that the proposed optimization can lead to significant capacity maximization for the secondary user as long as the primary user can afford.

## 1. Introduction

The cognitive radio (CR) principle has introduced the idea to exploit spectrum holes (i.e. bands) which result from the proven underutilization of the electromagnetic spectrum by modern wireless communication and broadcasting technologies [1]. The exploitation of these holes can be accomplished by the notion of cognitive radio networks (CRNs). A hierarchical model of a CRN consists of a primary (i.e. licensed) network (PN) and a secondary network (SN) where a secondary user (SU) exploits the available spectrum bands of the PN [2]. The objective in CRNs is to optimize the performance e.g. the capacity of the SU without causing harmful effects to the PU. In a sensing-based CRN, the power control (PoC) and the spectrum sensing (SpSe) at the SU are properly incorporated for the capacity optimization at the SU providing also the PU's protection [3]. To this end, in one hand, PoC at the SU with a constraint on the power interference caused at the PU gives access to the bands of the PN providing also the PU's protection [4][5]. On the other hand, SpSe at the SU with a constraint on the sensing threshold specifies the probability of the PU's detection and thus the PU's protection is provided either [2]. The SU's capacity optimization over the transmit power is studied in details for the CRNs [4][5] as well as the conventional wireless networks [12]. However, the SU's capacity optimization over the sensing threshold has not been recognized and studied. In [6], the authors have pointed out the importance of sensing threshold in CRNs; however, they do not provide any details on how the SU's capacity can be optimized over the sensing threshold. Hence, in this paper, the capacity optimization over the sensing threshold for CRNs is formulated and solved. The problem formulation results in a convex optimization problem where the concavity of the SU's capacity on sensing threshold is proved and the problem is solved using a Lagrange duality decomposition method in conjunction with a subgradient iterative algorithm [10][11][13][15].

The rest of this paper is organized as follows. Section 2 provides the system model of the sensing-based CRN with details on SpSe model. In section 3, we first formulate the convex optimization problem and in sequel we prove the concavity of the SU's capacity on the sensing threshold. We next provide the solution of this problem. Section 4 discusses the obtained numerical result that shows the achievable SU's capacity maximization and we conclude this work with section 5.

# 2. System model

### 2.1 Cognitive radio network model

We consider a sensing-based SS CRN with one PN and one SN which one provides a primary and secondary link respectively (Fig. 1). Both links consist of a transmitter and a receiver where for the PN are denoted as PU-Tx and PU-Rx and as SU-Tx and SU-Rx for the SN respectively. The links assumed to be flat fading channels (i.e. all frequency components of the signal will experience the same magnitude of fading) with additive white Gaussian noise (AWGN) [12]. The independent random variables of the AWGN are denoted with $n_p$ and $n_s$ for the primary and secondary link respectively which are assumed with mean zero and variance $N_0$. The PoC at the SU's transmitter (SU-Tx) aiming to protect the PU's receiver (PU-Rx) and for this reason the transmit power $P_t$ is applied as $P_t^0$ when the PU is idle or as $P_t^1$ when the PU is active with the following rule $P_t^0 > P_t^1$ [3]. Furthermore, perfect channel state information (CSI) is assumed to be available at the SU-Tx from the SU's receiver (SU-Rx) through a feedback channel.

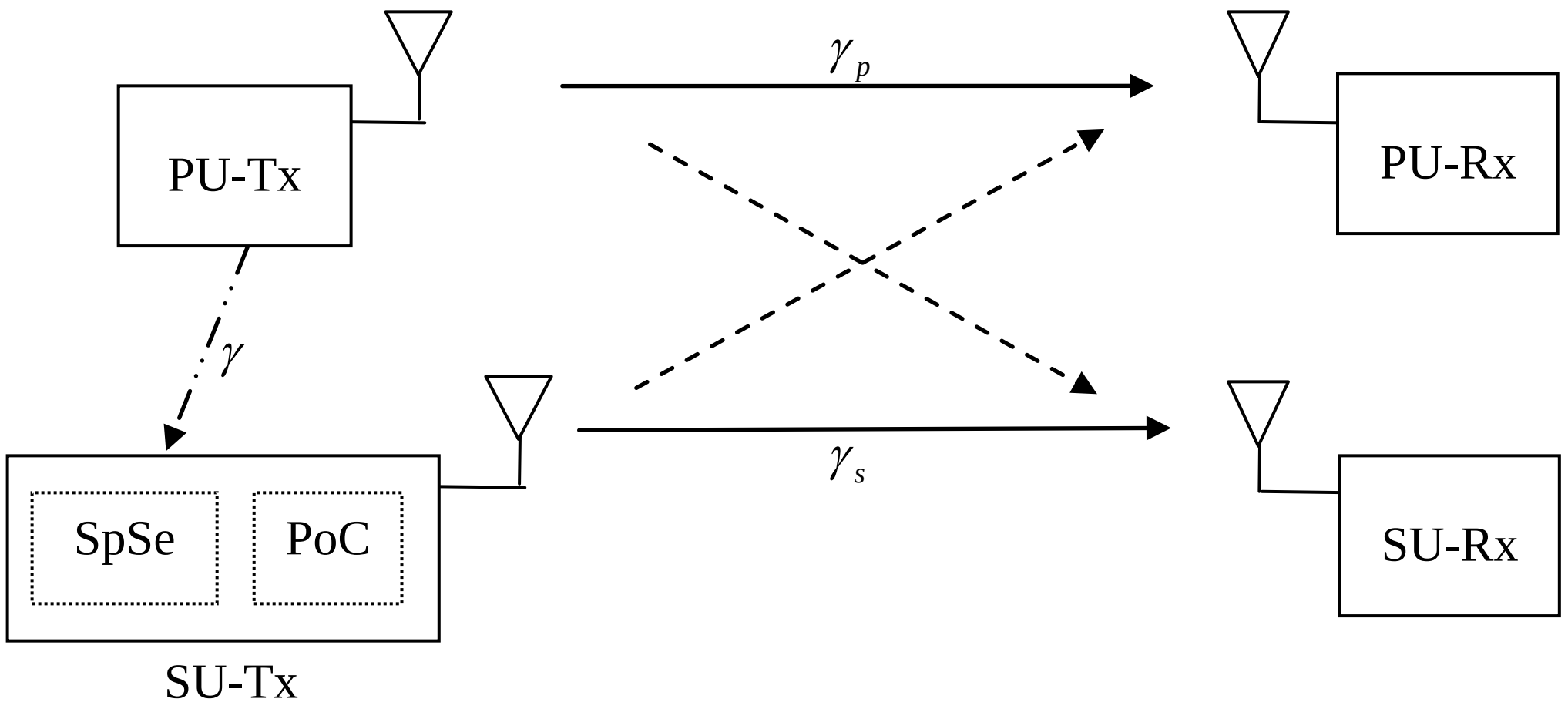

**Fig. 1** Cognitive Radio Network (CRN) model

### 2.2 Spectrum sensing model

The PU's activity is identified via a spectrum sensor that is employed at the SU-Tx. We assume an energy detector for SpSe which is able to sense the signal-to-noise-ratio (SNR) $\gamma$ for a specific time interval $\tau$ within the sampling frequency $f_s$ [7]. Spectrum sensing (SpSe) indicates whether the PU is active or idle by comparing the sensed SNR $\gamma$ with a sensing threshold $\eta$. The SpSe results in detection, missed detection, false alarm and no false alarm with probabilities $P_d$, $(1-P_d)$, $P_f$, $(1-P_f)$ respectively. The probabilities of false alarm and detection are defined as follows in relation with the hypotheses that the PU is idle or active denoted as $h_0$ and $h_1$ respectively

$$\begin{aligned} P_f &= \Pr[\gamma > \eta | h_0] \\ P_d &= \Pr[\gamma > \eta | h_1] \end{aligned} \tag{1}$$

Throughout this paper, we suppose a SpSe model with circularly Gaussian noise with mean zero and variance $\sigma^2$. Then the corresponding probabilities of false alarm and detection are defined as follows [7]

$$P_d(\eta, N) = Q\left(\left(\frac{\eta}{\sigma^2} - \gamma - 1\right)\sqrt{\frac{N}{2\gamma+1}}\right) \tag{2}$$

$$P_f(\eta, N) = Q\left(\left(\frac{\eta}{\sigma^2} - 1\right)\sqrt{N}\right) \tag{3}$$

where $Q(\cdot)$ is the complementary distribution function of the standard Gaussian[1] and $N$ is the number of samples taken from SpSe that is equal to $N = \tau f_s$.

Fig. 2 illustrates the complementary ROC (Receiver Operating Characteristic) of the considered SpSe model which plots the probability of missed detection $P_m = 1 - P_d$ vs. the probability of false alarm $P_f$ for different values of sensing threshold $\eta$. We depict the results obtained for different number of samples $N$ and variance equal to $\sigma^2 = 1$. The lines without marker are obtained for $\gamma = -15db$ and the lines with circle marker illustrate the results obtained with $\gamma = -12db$. Obviously, the SpSe behaves totally different by changing the sensing threshold $\eta$ and the number of the sensed samples $N$ for a given sensed SNR $\gamma$. Therefore, in order to retain a specific behavior for the SpSe mechanism in a sensing-based CRN, a proper sensing threshold value $\eta$ and number of samples $N$ for a given sensed SNR $\gamma$ must be selected. In [3], the authors have proposed the throughput optimization over sensing time $\tau$ when frame duration $T$ is considered and thus an optimal value of number of samples $N$ is obtained. In this paper, we investigate the capacity optimization over sensing threshold $\eta$ which is more generic and can lead to substantial capacity maximization regardless of the frame duration.

---

[1] $Q(x) = \frac{1}{\sqrt{2\pi}} \int_x^{\infty} e^{\left(-\frac{u^2}{2}\right)} du$

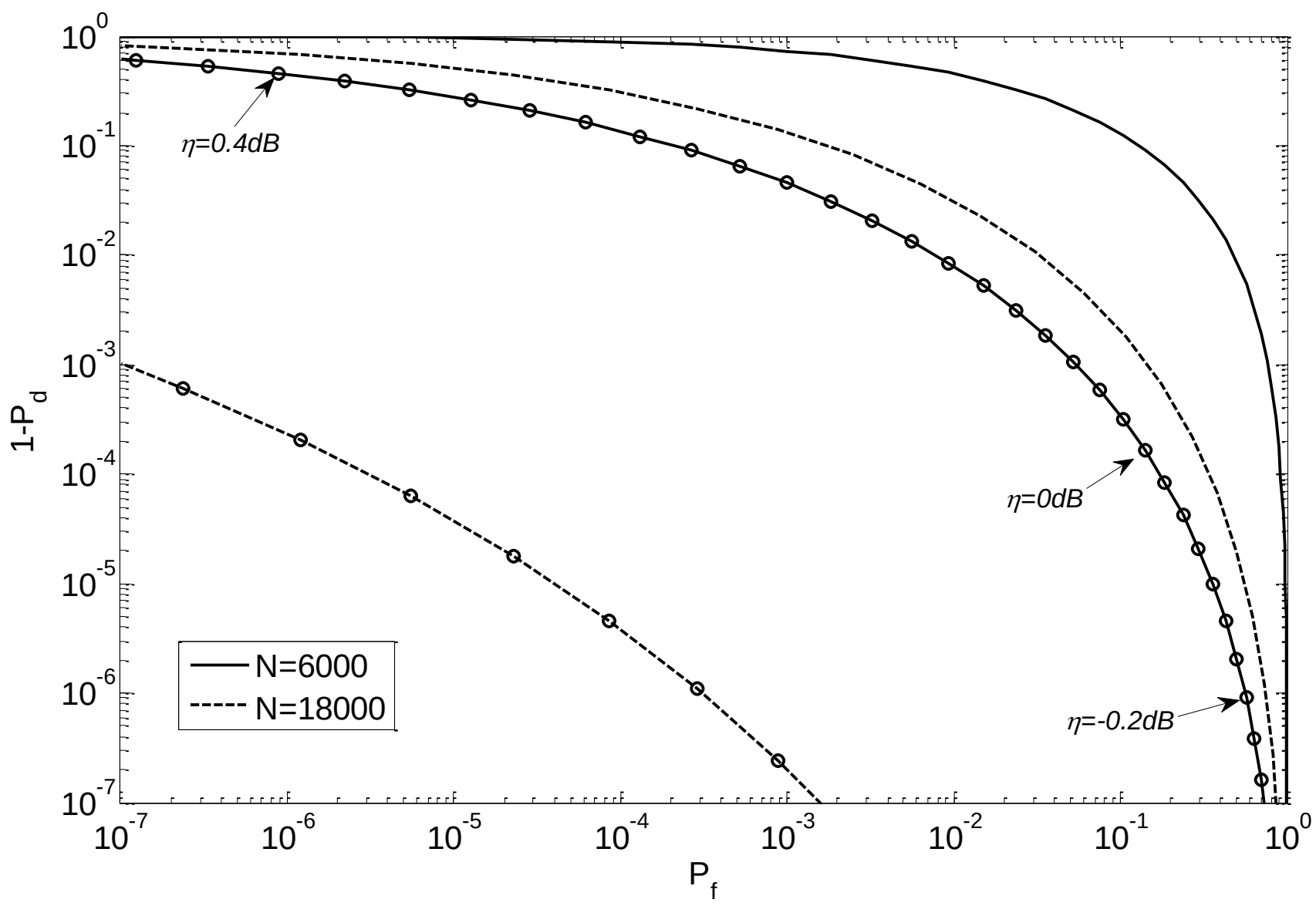

**Fig. 2** Operating characteristics of spectrum sensing (SpSe) for different values of sensing threshold $\eta$, number of sensed samples $N$ and sensed SNRs $\gamma$ i.e. $\gamma = -12dB$ (circle marker) and $\gamma = -15dB$ (no marker).

# 3. Capacity optimization over sensing threshold

In this section, we present the capacity optimization over the sensing threshold subject to the constraints for the PoC and SpSe in order to protect the PU. The formulated problem is a convex optimization problem and hence the concavity of SU's capacity on sensing threshold is proved. The problem's solution is provided by a Lagrange duality decomposition method in conjunction with a subgradient iterative algorithm.

## 3.1 Optimization problem formulation

Based on the probabilities that the PU is idle or active denoted as $\pi_0$ and $\pi_1$ respectively and the four results of SpSe, the following possible transmission scenarios are identified for the SU with the corresponding capacities: *the PU is idle with no false alarm denoted as* $C_0\left(P_t^0\right)\pi_0\left(1-P_f(\eta)\right)$, *the PU is idle with false alarm denoted as* $C_0\left(P_t^0\right)\pi_0 P_f(\eta)$, *the PU is active with detection denoted as* $C_1\left(P_t^1\right)\pi_1 P_d(\eta)$ *and the PU is active with missed detection denoted as* $C_0\left(P_t^0\right)\pi_1\left(1-P_d(\eta)\right)$. Thus, the overall SU's capacity is obtained as follows [3]

$$C_s\left(P_t^0, P_t^1, \eta\right) = C_0\left(P_t^0\right)\pi_0\left(1-P_f(\eta)\right) + C_1\left(P_t^1\right)\pi_0 P_f(\eta) + C_1\left(P_t^1\right)\pi_1 P_d(\eta) + C_0\left(P_t^0\right)\pi_1\left(1-P_d(\eta)\right) \quad (4)$$

The expression given in equation (4) is the objective function that we will next maximize over the transmit powers $P_t^0$ and $P_t^1$ and the sensing threshold $\eta$ for a sensed SNR $\gamma$ and a number of sensed samples $N$. The constraints on the transmit powers should regulate the average transmit power at the SU-Tx and the interference power at the PU-Rx that we express with the following equations denoted as $H\left(P_t^0, P_t^1, \eta\right)$ and $I\left(P_t^0, P_t^1, \eta\right)$ respectively

$$H\left(P_t^0, P_t^1, \eta\right) = \pi_0 E\left(P_t^0\right)\left(1 - P_f(\eta)\right) + \pi_0 E\left(P_t^1\right) P_f(\eta) + \pi_1 E\left(P_t^1\right) P_d(\eta) + \pi_1 E\left(P_t^0\right)\left(1 - P_d(\eta)\right) \quad (5)$$

$$I\left(P_t^0, P_t^1, \eta\right) = G_{sp}\pi_0 P_t^0\left(1 - P_f(\eta)\right) + G_{sp}\pi_0 P_t^1 P_f(\eta) + G_{sp}\pi_1 P_t^0 P_d(\eta) + G_{sp}\pi_1 P_t^1\left(1 - P_d(\eta)\right) \quad (6)$$

where $G_{sp}$ is the channel gain at the link between the SU-Tx and PU-Rx and $E(\cdot)$ is the expectation over the probability density function (pdf) $p(\gamma_s)$ of the fading channel at the SU link[2]. The constraint on the sensing threshold should regulate the probabilities of detection and missed detection, false alarm and no false alarm and henceforth a target probability of detection has been used so far [2][3]. However, in this work we choose to regulate the sensing threshold by using a more factual parameter than the target probability of detection. To this end, we define a level of the PU's capacity loss $C_{p,loss}$ that the PU can afford and thus the PU's channel state information (CSI) is involved [8].

Assuming that $P_{av}$ is the maximum average transmit power at the SU-Tx, $I_{pk}$ is the maximum peak interference power constraint that the PU-Rx can tolerate and that the PU's capacity loss $C_{p,loss}$ is less than some prescribed percentage $q$ over the maximum PU's capacity $C_{p,\max}$, then the SU's capacity optimization is formed as follows

$$\underset{\{P_t^0, P_t^1, \eta\}}{\max imize} \quad C_s\left(P_t^0, P_t^1, \eta\right) \quad (7)$$

$$\text{subject to} \quad H\left(P_t^0, P_t^1, \eta\right) \le P_{av} \quad (8)$$

$$I\left(P_t^0, P_t^1, \eta\right) \le I_{pk}$$

$$C_{p,loss} \le qC_{p,\max}$$

$$\text{with} \quad P_t^0 \ge 0,\ P_t^1 \ge 0,\ \eta \in [0, \infty]$$

for a constant $N \in [0, Tf_s]$, where $T$ is the frame duration and constant probabilities $\pi_0$ and $\pi_1$.

### 3.2 Concavity on sensing threshold

It is not difficult to observe that the problem is a convex optimization problem with respect to transmit powers $P_t^0$ and $P_t^1$. However, it is unclear whether this problem is a convex optimization problem with respect to sensing threshold $\eta$. In the following proposition, we show that the SU's capacity $C_s$ is concave on sensing threshold $\eta$.

*Proposition 1:* For the range of $\eta$ such that $P_d(\eta)$ is increasing and concave on $\eta$ and $P_f(\eta)$ is increasing and concave on $\eta$, the capacity $C_s$ is concave on $\eta$.

*Proof:* Differentiating both $P_f(\eta)$ and $P_d(\eta)$ with respect to $\eta$ gives:

---

[2] $E(P_t) = \int_{-\infty}^{+\infty} P_t\, p(\gamma_s)\, d\gamma_s$

$$P_f^{'}(\eta) = \frac{dP_f(\eta)}{d\eta} = \frac{1}{\sqrt{2\pi}}\left(\frac{\sqrt{N}}{\sigma^2}\right)\exp\left(-\left(\left(\frac{\eta}{\sigma^2}-1\right)\sqrt{N}\right)^2 \Big/ 2\right) \tag{9}$$

$$P_d^{'}(\eta) = \frac{dP_d(\eta)}{d\eta} = \frac{1}{\sqrt{2\pi}}\frac{1}{\sigma^2}\left(\sqrt{\frac{N}{2\gamma+1}}\right)\exp\left(-\left(\left(\frac{\eta}{\sigma^2}-\gamma-1\right)\sqrt{\frac{N}{2\gamma+1}}\right)^2 \Big/ 2\right) \tag{10}$$

Fig. 5 depicts the first derivatives $P_f^{'}(\eta)$ and $P_d^{'}(\eta)$ assuming $\gamma = -15dB$, $\sigma = 1$, $f_s = 6MHz$ and $\tau = 2ms$. The question we need to answer is if the first derivative of a function is increasing or decreasing or staying constant on the parameter of interest. Obviously, for $\eta > 0$, it is clear that $P_f^{'}(\eta) > 0$ and $P_d^{'}(\eta) > 0$. Thus, it can be said that both $P_f$ and $P_d$ are concave on $\eta$. However, there are critical values that could be identified as local extreme values i.e. local minima and local maxima [10]. These values are related with term $\left|\eta/\sigma^2 - 1\right|$ for the probability of false alarm $P_f$ and with term $\left|\eta/\sigma^2 - \gamma - 1\right|$ for the probability of detection $P_d$. Fig.6 shows these values $\left|\eta/\sigma^2 - 1\right|$ and $\left|\eta/\sigma^2 - \gamma - 1\right|$ where the concavity of $P_d$ and $P_f$ on $\eta$ is proved.

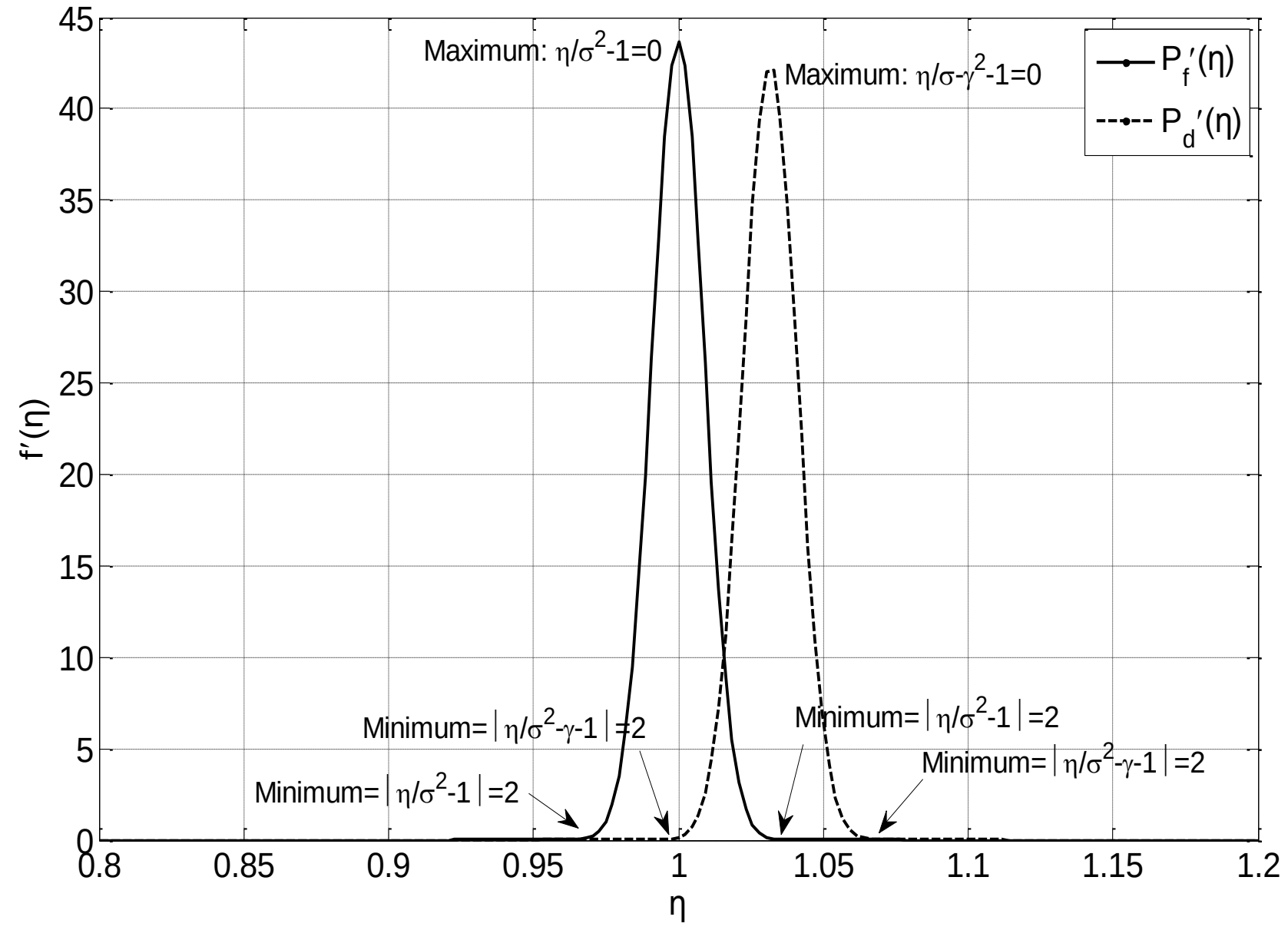

**Fig. 3** Local minima and maxima of the first derivative $P_f^{'}(\eta)$ and $P_d^{'}(\eta)$

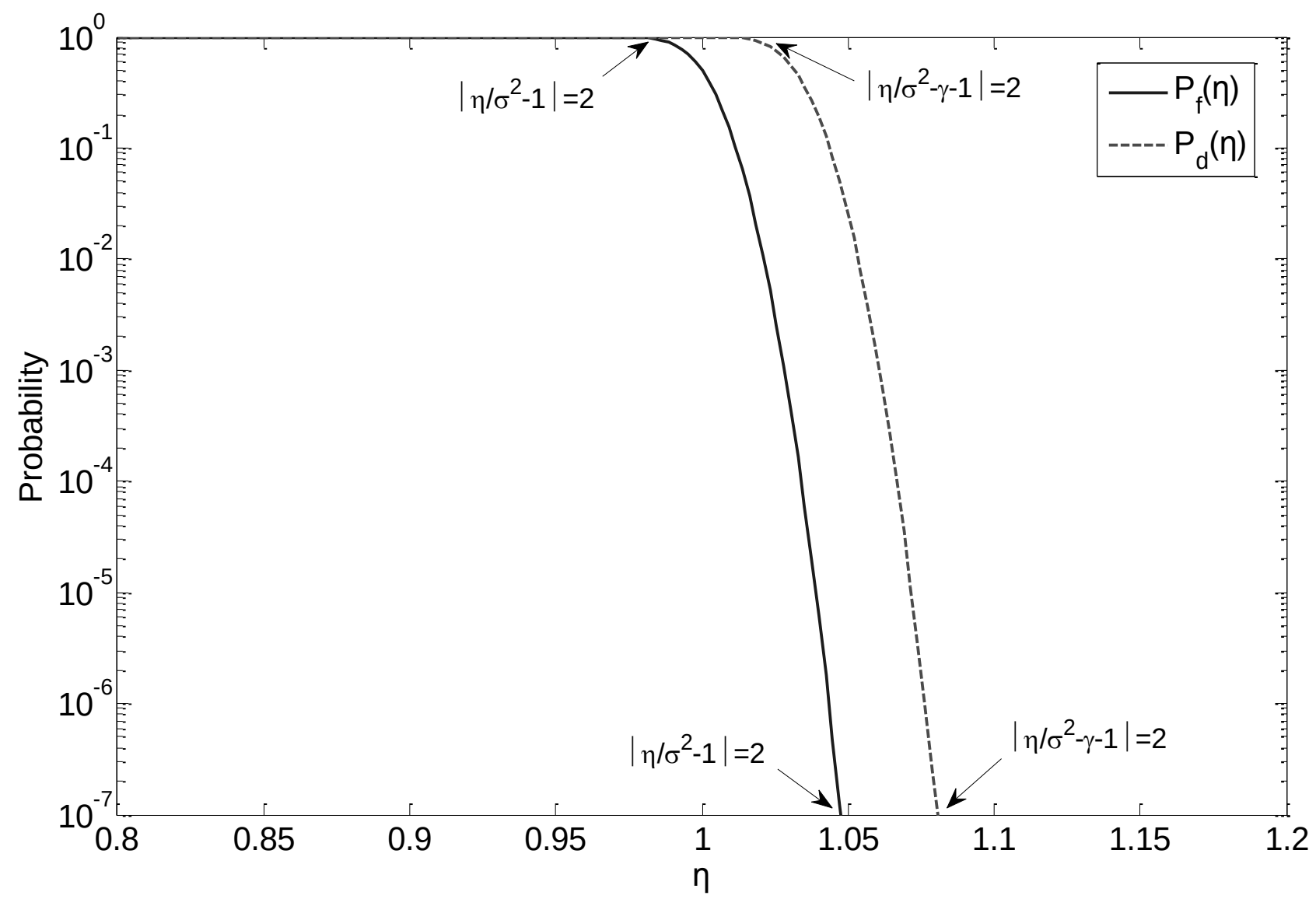

**Fig. 4** Probability of false alarm $P_f$ and detection $P_d$ vs. sensing threshold $\eta$

We now check the concavity of SU's capacity $C_s$ and thus we derive the first derivative of SU's capacity $C_s$ with respect to $\eta$

$$C_s^{'}(\eta) = \frac{dC_s(\eta)}{d\eta} = [C_1\pi_0 - C_0\pi_0]\frac{dP_f}{d\eta} + [C_1\pi_1 - C_0\pi_1]\frac{dP_d}{d\eta} \quad (11)$$

and then

$$C_s^{'}(\eta) = \pi_0 P_f^{'}(\eta)(C_1 - C_0) + \pi_1 P_d^{'}(\eta)(C_1 - C_0) \quad (12)$$

Since both $P_f^{'}(\eta)$ and $P_d^{'}(\eta)$ are concave on $\eta$ and $(C_1 - C_0) < 0$ holds, the first derivative of capacity is decreasing on $\eta$ i.e. $C_s^{'}(\eta) < 0$. This implies that the SU's capacity $C_s$ is concave on sensing threshold $\eta$ and thus the optimization problem over sensing threshold $\eta$ is concave either.

### 3.3 Lagrange duality and subgradient iterative algorithm

In order to solve the convex optimization problem defined in equation (4), a dual decomposition method can be applied for a value for the sensing threshold $\eta$ [15]. The *Lagrangian* of (4) is defined as

$$L(P_t^0, P_t^1, \lambda) = L(P_t^0, P_t^1, \eta) - \lambda(H(P_t^0, P_t^1, \eta) - P_{av}) \quad (13)$$

and the *Lagrangian dual function* is defined as

$$q(\lambda) = \sup_{\{P_t^0, P_t^1\}} \{L(P_t^0, P_t^1, \lambda) | I(P_t^0, P_t^1, \eta) \leq I_{pk}\} \quad (14)$$

The dual function can then be minimized to obtain an upper bound on the *optimal value* $C_s^*$ of the optimization problem in (4)

$$\underset{\lambda}{\text{minimize}} \quad q(\lambda) \tag{15}$$

where the *optimal dual objective* $q^*$ forms the *duality gap* $C_s^* - q^*$ which is indeed zero since the Karush–Kuhn–Tucker (KKT) conditions are satisfied.
Based on the above problem formulation, the optimal solution of the transmit power is obtained as follows [4]

$$P_t^0 = \left(\frac{1}{\lambda} - \frac{N_0}{\gamma_s}\right)^+, \quad \text{for} \quad \gamma_s \geq \lambda \tag{16}$$

$$P_t^1 = \begin{cases} \left(\frac{1}{\lambda} - \frac{N_0}{\gamma_s}\right)^+ & \text{for} \quad \gamma_s \geq \lambda \;\; and \;\; G_{sp} < I_{pk}/(1/\lambda - N_0/\gamma_s) \\ \frac{I_{pk}}{G_{sp}} & \quad\quad\; \gamma_s \geq \lambda \;\; and \;\; G_{sp} \geq I_{pk}/(1/\lambda - N_0/\gamma_s) \end{cases} \tag{17}$$

where the following is applied $P_t^0 = 0$ and $P_t^1 = 0$ for $\gamma_s < \lambda$.

After the decomposition performed above, we will use a subgradient iterative algorithm for solving the *nondifferentiable* function of $C_s$ in (4) [11]. The subgradient of $q(\lambda)$ is given by the following proposition.
*Proposition 2:* The subgradient of $q(\lambda)$ is $g(P_t^0, P_t^1) = P_{av} - H(P_t^0, P_t^1)$ for the *ith* iteration (i.e. a given sensing threshold $\eta$) and the $g(\{P_t^0, P_t^1\}_\lambda)$ is an element of $\partial q(\lambda)$[3].
*Proof:* For any $\mu \in dom(q)$, since $q(\mu)$ is obtained by maximizing $L(P_t^0, P_t^1, \mu)$ over $P_t^0, P_t^1 \in dom(C_s)$, we have $q(\mu) \geq L(\{P_t^0, P_t^1\}_\lambda, \mu)$[4] [13]. Moreover, since $\{P_t^0, P_t^1\}_\lambda$ achieves the maximum, we have $q(\lambda) = L(\{P_t^0, P_t^1\}_\lambda, \mu)$. Combining the pieces, we obtain

$$\begin{aligned} q(\mu) &\geq L(\{P_t^0, P_t^1\}_\lambda, \mu) \\ &= L(\{P_t^0, P_t^1\}_\lambda, \lambda) + \left[L(\{P_t^0, P_t^1\}_\lambda, \mu) - L(\{P_t^0, P_t^1\}_\lambda, \lambda)\right] \\ &= q(\lambda) + \mathrm{H}(\{P_t^0, P_t^1\}_\lambda)^T (\mu - \lambda) \end{aligned} \tag{18}$$

Afterwards, the problem can be solved by the following Algorithm, which requires the calculation of the subgradient $g$ at each iteration.

***Algorithm****: Subgradient iterative algorithm*

Parameters: constant step size $\alpha$ and constant convergence value $\in$

- Initialize: variable $\lambda^{(k)} = 1$ and counter $k = 1$, where $\lambda$ is the SNR at the SU-Rx $\gamma_s$

1. For a sensing threshold $\eta$ (iteration).
2. For a specific $P_{av}$, calculate the expectation of transmit powers $E(P_t^0)$ and $E(P_t^1)$
3. Calculate the subgradient $g$ as follows
$$g(P_t^0, P_t^1, \eta) = P_{av} - H(P_t^0, P_t^1, \eta)$$
4. a) Find the common optimum $\gamma_s^*$ iteratively from the $\lambda^{(k+1)} = \lambda^{(k)} - \alpha \cdot g^{(k)}$ b) calculate iteratively until the convergence rule is reached.

---

[3] $\partial q(\lambda)$ denotes the set of all sub gradients at $\lambda$ that is called the *subdifferential*.

[4] where $\{P_t^0, P_t^1\}_\lambda$ it the pair values of the transmit powers $P_t^0$ and $P_t^1$ for the Lagrange multiplier value $\lambda$.

5. Exit from the algorithm and calculate the corresponding capacity from (4) for the optimal value pair $\left[\gamma_s^*, \eta^*\right]$, where $\eta^*$ satisfies the $C_{p,loss}$.

In the subgradient iterative algorithm described above, the sensing threshold $\eta$ is matched in a specific capacity loss $C_{p,loss}$ using the separation principle of wireless networking [9][14]. Given this principle, a probability of missed detection $P_m = 1 - P_d$ at the SU-Tx will result in an outage probability for the PU-Rx that presents the probability that the transmission is decoded with a large error probability at the PU-Rx [12]. In particular, if the received SNR at the PU-Rx is below $\gamma_{p,\min}$ then the bursty transmitted bits are decoded with an error probability approaching one, and thus the receiver PU-Rx declares an outage with the following probability

$$P_{out} = \Pr\left[\gamma_p < \gamma_{p,\min}\right] \tag{19}$$

In this case, the capacity loss at the PU's link is obtained as follows

$$C_{p,loss} = C_{p,\max} - \int_{\gamma_{p,\min}}^{\infty} \log_2(\gamma_p) p(\gamma_p) d\gamma_p \tag{20}$$

where $p(\gamma_p)$ is the probability density function (pdf) of the fading distribution at the PU's link, while the maximum achievable capacity at the PU's link is equal to $C_{p,\max} = \int_0^{\infty} \log_2(\gamma_p) p(\gamma_p) d\gamma_p$. Thus, a capacity loss $C_{p,loss}$ which depends on the outage probability $P_{out}$ will dictate the missed detection probability $P_m$ that matches in a sensing threshold $\eta$ value.

# 4. Numerical results

Fig. 3 depicts the results obtained for SU's capacity maximization versus sensing threshold $\eta$ for different values of sensed SNR $\gamma$ at the spectrum sensor of the SU-Tx and average transmit power $Pav$ of the SU-Tx at the SU's link. We assume a number of samples equal to $N = 12000$ that means a sensing time equal to $\tau = 2ms$ for a sampling frequency $f_s = 6MHz$ and variance equal to $\sigma^2 = 1$. The optimal power allocation at the SU's link is taking place over a Rayleigh fading channel with unit mean and AWGN with variance $N_0 = 1$ [5]. Besides, we assume an interference power constraint equal to $I_{pk} = 0db$ while the PU's activity is considered as $\pi_1 = 0.4$. It is observed that a proper power allocation and sensing threshold adaptation for a given sensed SNR $\gamma$ results in significant capacity increase for the SU especially for large values of the average transmit power e.g. $Pav = 15dB$. This maximization is getting lower for lower values of $Pav$ e.g. $Pav = 5dB$ and becomes negligible when $Pav$ is lower than the considered peak interference power constraint $I_{pk}$ i.e. when $Pav < I_{pk}$. This is due to the fact that a missed detection do not affect the system's behavior since the transmit powers are now equal i.e. $P_t^0 = P_t^1$. Moreover, low values of sensed SNR $\gamma$ will lead to further capacity maximization for a given sensing threshold $\eta$.

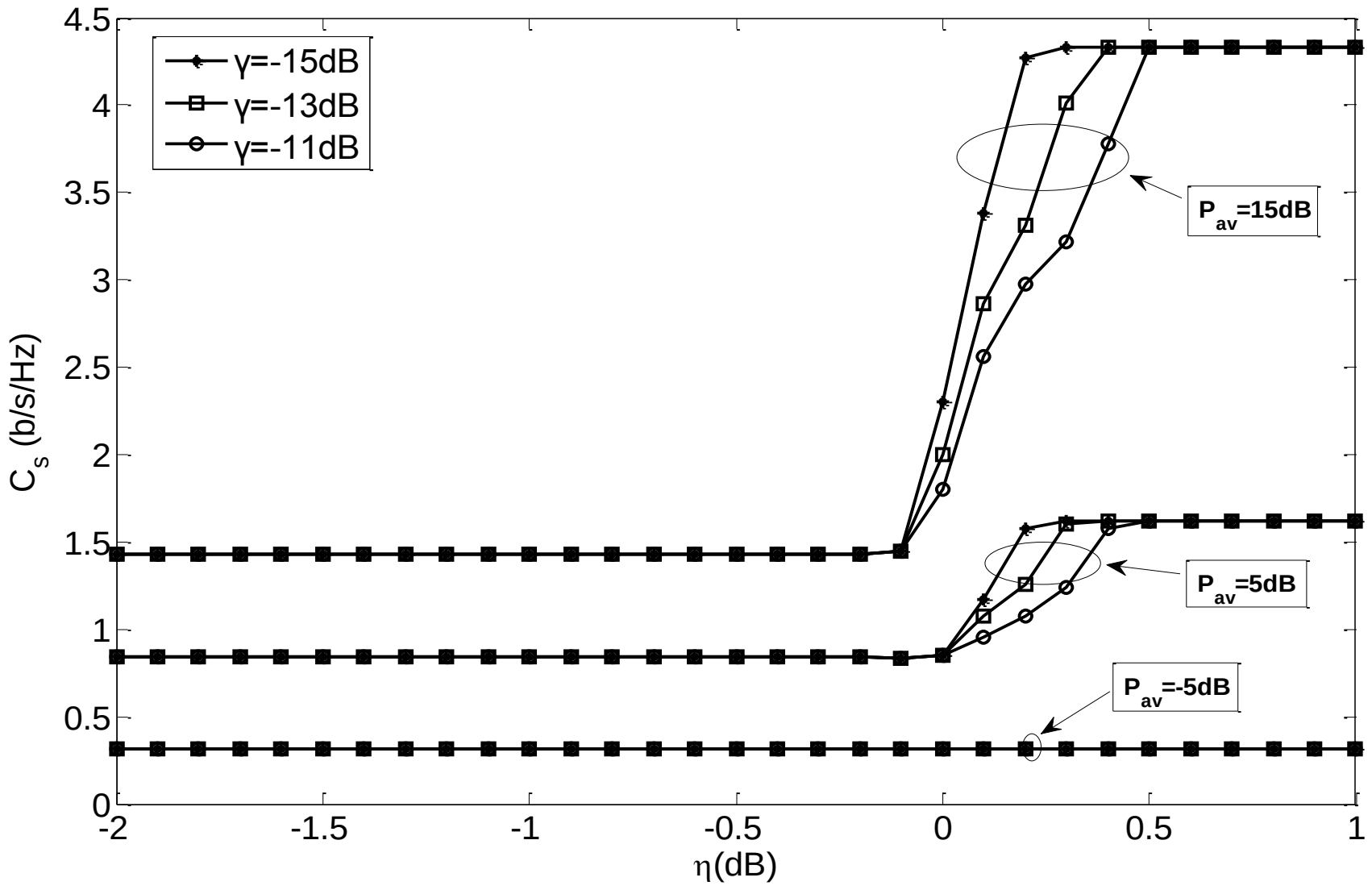

**Fig. 5** SU's capacity $C_s$ vs. sensing threshold $\eta$

Fig. 4 illustrates the PU's capacity loss $C_{p,loss}$ versus the sensing threshold $\eta$. The CRN setup is the previous one in terms of the implementation details of the PoC and the SpSe at the SU-Tx. We depict the results for different average transmit powers $Pav$ of the PoC and sensed SNRs $\gamma$ at the SpSe of the SU-Tx either. The results are obtained assuming that the probability of missed detection $P_m$ yields an outage probability i.e. $P_m = P_{out}$. Thus, a specific sensing threshold $\eta$ value represents a specific missed detection probability $P_m$ which is matched next into a CSI at the Pu-Rx $\gamma_p$ that yields an outage. Fig. 3 shows that the higher the sensing threshold $\eta$, the higher the capacity loss $C_{p,loss}$ is become. This is true since the probability of missed detection $P_m$ is getting higher. The probability of missed detection $P_m$ leads to lower values in CSI at the Pu-Rx $\gamma_p$ and thereafter in lower achievable capacities at the PU i.e. higher capacity loss $C_{p,loss}$. Obviously, there exists a fundamental tradeoff between the achievable capacity maximization and the affordable capacity loss $C_{p,loss}$ at the PU-Rx when different sensing threshold values are considered. Hence, the PU's capacity loss $C_{p,loss}$ can act as a factual quality of service metric that can be satisfied by properly adapting the sensing threshold. Finally, the lower the sensing threshold is become the lower the sensed SNR $\gamma$ that brings about capacity loss maximization.

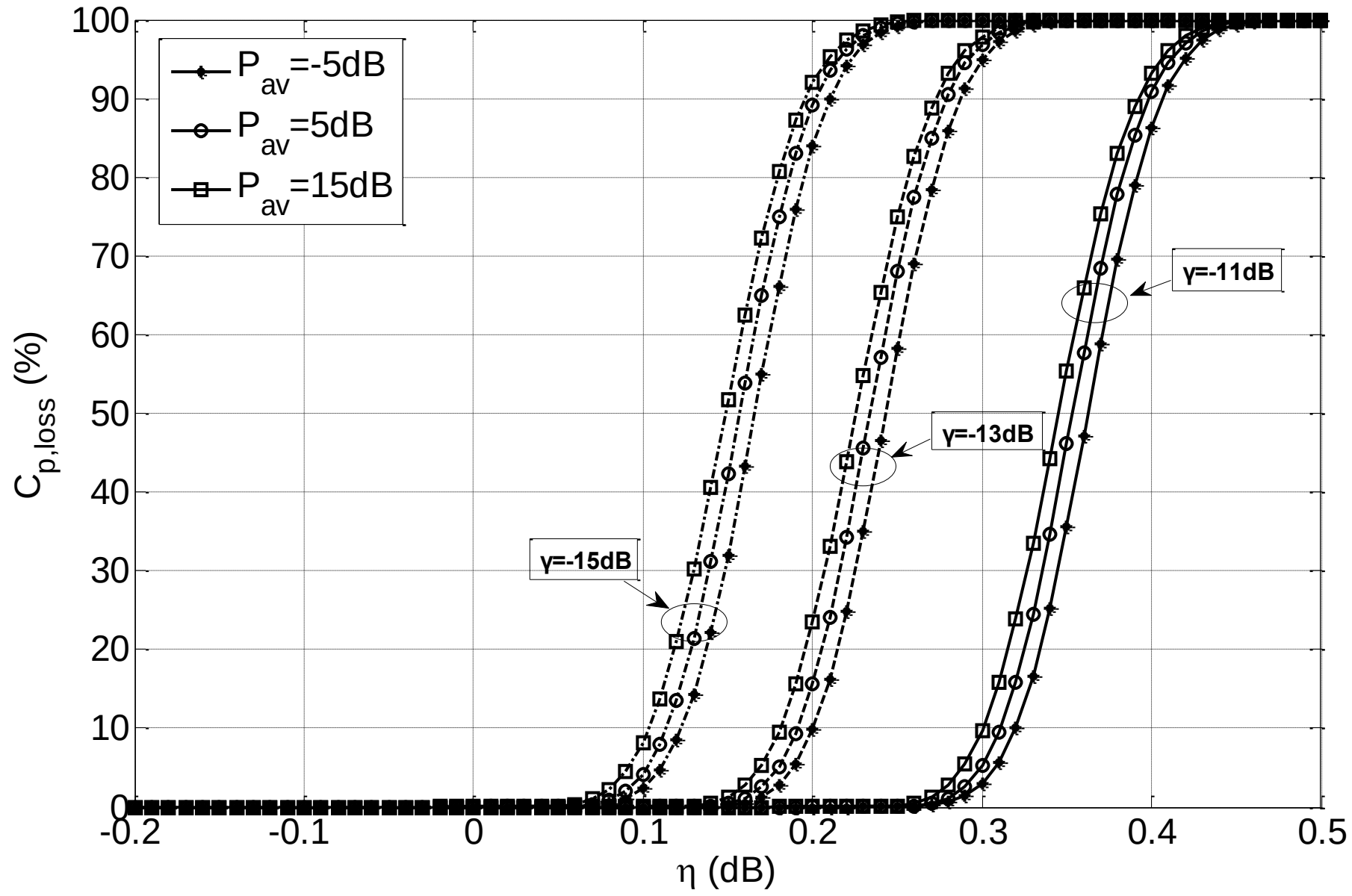

**Fig. 6** PU's capacity loss $C_{p,loss}$ vs. sensing threshold $\eta$

# 5. Conclusion

In this paper, we proposed the capacity optimization over sensing threshold for sensing-based cognitive radio networks (CRNs). In particular, we consider a sensing-based spectrum sharing CRN in which both power control (PoC) and spectrum sensing (SpSe) are employed for the PU's protection. Assuming the constraints for both the PoC and the SpSe, the proposed optimization is being a convex optimization problem that is solved using a Lagrange duality decomposition method. In sequel, a subgradient iterative algorithm provides the optimum values for both the transmit power and the sensing threshold of the PoC and SpSe respectively. The numerical results show that the maximization of the SU's capacity is sufficiently large when it is not harmful for the PU's transmission that can be controlled using a constraint on the PU's capacity loss, which represents the actual state of the primary link.